\documentclass[a4paper, amsfonts, amssymb, amsmath, reprint, nofootinbib,superscriptaddress,twoside, english,onecolumn, pra]{revtex4}

\usepackage[english]{babel}
\usepackage[utf8]{inputenc}
\DeclareUnicodeCharacter{2009}{\,} 
\usepackage[colorinlistoftodos, color=green!40, prependcaption]{todonotes}
\usepackage{lipsum}

\usepackage[pdftex, pdftitle={Article}, pdfauthor={Author}]{hyperref} %

\usepackage[normalem]{ulem}

\definecolor{FF}{RGB}{14, 159, 14}%

\usepackage{graphicx}
\usepackage{subfigure}
\usepackage{tabulary}
\usepackage{amsmath}
\usepackage{color} %
\usepackage{float}
\usepackage{amsfonts}
\usepackage{amssymb}
\usepackage{mathrsfs}
\usepackage{mathtools}
\usepackage{dsfont}
\usepackage{eucal}
\usepackage{braket}
\usepackage[left=2cm,right=2cm,top=3cm,bottom=2.5cm]{geometry}
\usepackage{hyperref}
\hypersetup{
    colorlinks,
    linkcolor={black},
    citecolor={black},
    urlcolor={black}
}
\usepackage{caption}
\usepackage{multirow}
\usepackage{enumitem}
\setlist[itemize]{leftmargin=*}

\begin{document}
\title{Control and Transient Spectroscopy of Engineered Spin-Cavity Back-Action}

\author{Fatemeh Fani Sani}
\affiliation{The Institute for Quantum Computing, University of Waterloo, Waterloo, Ontario, N2L 3G1, Canada}
\affiliation{Department of Physics and Astronomy, University of Waterloo, Waterloo, Ontario, N2L 3G1, Canada}

\author{George Nichols}
\affiliation{The Institute for Quantum Computing, University of Waterloo, Waterloo, Ontario, N2L 3G1, Canada}

\author{Ivar Taminiau}
\affiliation{The Institute for Quantum Computing, University of Waterloo, Waterloo, Ontario, N2L 3G1, Canada}

\author{Saba Sadeghi}
\affiliation{The Institute for Quantum Computing, University of Waterloo, Waterloo, Ontario, N2L 3G1, Canada}

\author{Hamid R. Mohebbi}
\affiliation{The Institute for Quantum Computing, University of Waterloo, Waterloo, Ontario, N2L 3G1, Canada}
\affiliation{High Q Technologies Inc, Waterloo, Ontario, N2L 3G1, Canada}

\author{David G. Cory}
\affiliation{The Institute for Quantum Computing, University of Waterloo, Waterloo, Ontario, N2L 3G1, Canada}
\affiliation{Department of Chemistry, University of Waterloo, Waterloo, Ontario, N2L 3G1, Canada}    

\author{Troy W. Borneman}
\email{troyborneman@gmail.com}
\affiliation{The Institute for Quantum Computing, University of Waterloo, Waterloo, Ontario, N2L 3G1, Canada}
\affiliation{High Q Technologies Inc, Waterloo, Ontario, N2L 3G1, Canada}

\date{\today} %

\begin{abstract}

\noindent We present an experimental arrangement that permits engineering of cavity back-action on a mesoscopic spin ensemble. 
By coupling a superconducting thin-film Nb microstrip resonator to a Trityl OX63 electron spin sample, we access different regimes of spin-cavity dynamics by designing the ensemble size, effective coupling strength, cavity temperature, and spin saturation. We performed transient spectroscopy measurements under continuous microwave drive in the strong radiation damping regime. These measurements exhibit a long-lived plateau response that distinguishes important features of spin-cavity models, such as the radiation damping Bloch equations and Maxwell-Bloch equations. We demonstrate control of the plateau response through adjustment of temperature, microwave drive power, and variable spin saturation. The presented experimental arrangement serves as a robust system to explore the space of spin-cavity dynamics and develop new quantum devices that harness the complexity of mesoscopic spin ensembles coherently interacting with high quality factor cavities.
\end{abstract}

\maketitle

\section{Introduction}

Increasing size and complexity in experimental quantum systems necessitates a deeper understanding of the dynamics between mesoscopic ensembles of two-level systems and quantum harmonic oscillators is necessary. We consider here a \emph{spin-cavity} system consisting of an electron spin ensemble coupled to a high quality factor superconducting resonator \cite{benningshof_superconducting_2013,morton_hybrid_2011}. A number of other systems implementing quantum information processing and sensing admit the same physical description, including superconducting qubits, trapped ions, and neutral atoms \cite{fink_dressed_2009,yang_probing_2020,wang_controllable_2020,xiang_hybrid_2013}. 

Performing useful operations requires a large number of two-level systems (spins), which becomes intractable for classical simulation. Experimental systems that provide characterization and control of large spin-cavity systems are a crucial tool for developing useful quantum devices.

An electron spin ensemble coupled to a superconducting resonator is a prototypical spin-cavity system that has been investigated for the purposes of quantum information processing \cite{wesenberg_quantum_2009,kurizki_quantum_2015,morton_hybrid_2011,ranjan_pulsed_2020}, quantum sensing
and implementations of quantum memories \cite{ranjan_multimode_2020,osullivan_random-access_2021,morton_storing_2018,grezes_multimode_2014}. Several regimes of spin-cavity dynamics using electron spin ensembles have been demonstrated utilized, including weak coupling (conventional spectroscopy) \cite{bushev_ultralow-power_2011,staudt_coupling_2012}, strong coupling with resolved Rabi splitting \cite{abe_electron_2011,dold_high-cooperativity_2019,schuster_high-cooperativity_2010}, and superradiant behavior \cite{angerer_superradiant_2018,rose_coherent_2017}. Each regime offers distinct advantages and disadvantages depending on the application. Consequently, the ability to tune parameters within an experiment significantly enhances the device's power. An important example is measurement back-action, which can be used to mediate long-range entanglement of disjoint spins \cite{mirkamali_entanglement_2018,mirkamali_mesoscopic_2019,astner_coherent_2017,borneman_parallel_2012,norcia_cavity-mediated_2018} and create correlated spin-cavity states that greatly enhance the sensitivity of quantum sensors \cite{munoz-arias_phase-space_2023,koppenhofer_dissipative_2022}. Measurement back-action has a long history of study in magnetic resonance, where radiation damping was noted early in the development of NMR \cite{bloembergen_radiation_1954,bloom_effects_1957}. 
Radiation damping is generally considered detrimental, introducing unwanted artifacts in two-dimensional correlation spectroscopy (COSY) measurements \cite{mccoy_threequantum_1990} and leading to incorrect characterization of sample T1 \cite{sodickson_initiation_1996}. The connection between radiation damping and superradiance has also been established, where a time-delayed hyperbolic secant superradiant burst is observed \cite{vlassenbroek_radiation_1995,dicke_coherence_1954}. Further studies of measurement back-action in mesoscopic spin ensembles have generally been limited in the context of engineering and control to considering a single spin with magnetic moment enhanced by the number of spins \cite{slichter_principles_1990}. 
However, the true state structure of mesoscopic ensembles is significantly more complex \cite{gunderman_lamb_2021,wesenberg_mixed_2002,temnov_superradiance_2005}, providing a powerful resource for advanced quantum devices if the dynamics can be understood, modeled, and controlled.

We present a new experimental arrangement and set of methodologies to study and control spin-cavity dynamics. We inductively coupled a 2.5 nL frozen glass solution of Trityl OX63 free-radical electron spins to a homogeneous microwave field generated by a $\lambda/2$ Niobium (Nb) superconducting microstrip resonator patterned on a sapphire chip. This arrangement was previously used to study cavity-induced spin linewidth narrowing \cite{akhmetzyanov_electron_2023} and allows us to tune device parameters, precisely characterize dynamics, and perform control sequences that explore control in the space of spin-cavity models. The response of our system to low-amplitude continuous microwave irradiation is studied using transient spectroscopy methods \cite{putz_protecting_2014,angerer_ultralong_2017}. We observe a unique long-lived plateau response that distinguishes different models of spin-cavity dynamics and leverages back-action as a quantum resource that enables detailed spin response studies for an important class of low-temperature samples with long $T_1$.

\section{Modeling Spin-Cavity Dynamics}
\subsection{Tavis-Cummings Model}

Assuming a Markov environment, the dynamics of a spin-cavity system can be fully described by a Lindblad master equation that includes both coherent Hamiltonian dynamics and dissipative processes:
\begin{equation}
    \dot {\rho} = -i[\mathcal{H}_\text{TC}, \rho] + \sum_j\mathcal{D_j}[\rho],
\end{equation}
The $\mathcal{H}_\text{TC}$ Hamiltonian is given by the Tavis-Cummings (TC) model \cite{tavis_exact_1968} under a rotating-wave approximation:
\begin{equation}
    \mathcal{H}_\text{TC} = \Delta_c a^\dagger a + \sum_i\frac{1}{2}\Delta_s^i\sigma_z^i + i\sum_i\frac{g_0^i}{2}(\sigma_-^ia^\dagger -\sigma_+^i a)+i\left(\mathcal{E}a^\dagger-\mathcal{E}^*a\right),
\end{equation}
where $a$ and $a^\dagger$ are the standard photon creation and annihilation operators for the cavity, respectively; $\sigma_\alpha^i$ are the standard spin-1/2 Pauli operators for the $i^\text{th}$ spin in the ensemble; $\Delta_c=\omega_c-\omega_r$ is the offset of the cavity resonance frequency, $\omega_c$, from the rotating frame frequency, $\omega_r$; $\Delta_s^i=\omega_0^i-\omega_r$ is the offset of the spin resonance frequency of the $i^\text{th}$ spin with gyromagnetic ratio $\gamma^i$, $\omega_0^i=\gamma^iB_0$, for a quantizing field of strength $B_0$; and $g_0^i = \mu_B/\hbar\sqrt{2\mu_0\hbar\omega_0^i/V_c}$ is the interaction strength of photon exchange between the $i^\text{th}$ spin the cavity mode of volume $V_c$.

The dissipators, $\mathcal{D_j}$, correspond to non-unitary relaxation of the spin operators under amplitude damping, $\mathcal{D}_1$ ($T_1$), and phase damping, $\mathcal{D}_2$ ($T_2$), and relaxation of the cavity operators under amplitude damping, $\mathcal{D}_c$ \cite{alsing_spontaneous_1991}:
\begin{align}
    \mathcal{D}_\text{spin,1} &= \gamma_1(\sigma_-\rho\sigma_+ - \frac{1}{2}(\sigma_+\sigma_-\rho + \rho\sigma_+\sigma_-)), \\
    \mathcal{D}_\text{spin,2} &= \gamma_2(\sigma_z\rho\sigma_z - \rho), \\
    \mathcal{D}_\text{cav} &= \kappa(2a\rho a^{\dagger} - a^\dagger a\rho-\rho a^\dagger a).
\end{align}
The cavity quality factor, $Q$, dictates the strength of the cavity dissipation, $\kappa = \omega_0/(Q)$, and the spin relaxation expressions, $T_1$ and $T_2$, dictate the spin dissipation, $\gamma_1=1/T_1$ and $\gamma_2=(2T_1-T_2)/(2T_1 T_2)$. Taken together, these three parameters can define controls to manipulate the undriven dynamics of the system. We have also introduced microwave irradiation as a cavity drive Hamiltonian of strength $|\mathcal{E}|$ that leads to an effective Rabi Hamiltonian that acts on the spins in the rotating frame, $H_\text{Rabi}=\omega_1(\cos{\phi}J_x+\sin{\phi J_y})$.

The structure of the TC Hamiltonian is complex and admits a direct sum representation over a set of collective angular momentum operators \cite{gunderman_lamb_2021}. In general, simulations of the TC Hamiltonian are limited to small spin ensembles and truncated cavity populations. Considering larger systems requires making approximations that take advantage of the permutation symmetry of the largest angular momentum subspace, the Dicke subspace \cite{dicke_coherence_1954,kirton_introduction_2019}. It has been shown that while the Dicke subspace is nearly unpopulated at normal experimental temperatures (even 10 mK), the maximally degenerate subspaces that contain most of the spin population lead to similar behavior \cite{gunderman_thermal_2024} and, under an assumption of few excitations in the system, the spin ensemble may be modeled as a collective system via a Holstein-Primakoff approximation \cite{holstein_field_1940,gyamfi_introduction_2019}, leading to the Dicke Hamiltonian \cite{garraway_dicke_2011,roses_dicke_2020}:
\begin{equation}
    \mathcal{H}_\text{D} = \Delta_c a^\dagger a + \frac{1}{2}\Delta_s J_z + i{g}(J_-a^\dagger - J_+ a)+i\left(\mathcal{E}a^\dagger-\mathcal{E}^*a\right),
\end{equation} 
where the spin operators are now collective angular momentum operators, $J_\alpha=1/\sqrt{N}\sum_i\sigma_\alpha^i$, and the strength of the photon exchange term is now given by a collective strength that depends on the size, $N$, and polarization, $p$, of the spin ensemble \cite{benningshof_superconducting_2013}:
\begin{equation}
    g = g_0\sqrt{pN}=\mu_B\sqrt{\frac{2\mu_0 \omega_0pN}{\hbar V_c}}.
\end{equation}

\subsection{Maxwell-Bloch Equations}
The Lindblad master equation may be recast as a set of coupled differential equations in the Schrodinger formalism corresponding to tracking the evolution of a closed set of observables, $\mathcal{O}$, leading to a set of Maxwell-Bloch equations \cite{zens_critical_2019}:
\begin{align}
    \langle \dot{a} \rangle &= -(\kappa + i\Delta_c)\langle a \rangle + g\langle J_-\rangle + |\mathcal{E}|, \\
    \langle \dot{J_-}\rangle &= -(\gamma_1/2 + \gamma_2+ i\Delta_s)\langle J_-\rangle + g\langle a J_z\rangle, \\
    \langle\dot{J_z}\rangle &= -\gamma_1(1+\langle J_z\rangle) - 2g(\langle a^\dagger J_-\rangle + \langle a J_+\rangle).
\end{align}
These equations are not closed in this form, as no equations of motion are defined for spin-cavity correlation terms, i.e. $\langle a\sigma_z\rangle$. 
The seminal Maxwell-Bloch equations invoke a semiclassical first-order mean-field theory approximation by dropping terms of order 2 or greater in a cumulant expansion of these terms \cite{kubo_generalized_1962} $\langle AB\rangle = \langle A\rangle\langle B\rangle + \langle AB\rangle_c \approx \langle A\rangle\langle B\rangle$.
Applying this approximation, assuming $\Delta_c=\Delta_s=0$, and defining real-valued Hermitean spin and cavity observables yields Maxwell-Bloch equations to first order, MBE1 \cite{wu_self-oscillation_2018}:
\begin{align}
    \langle \dot{E} \rangle &= -\kappa\langle E \rangle + \frac{g}{2}\langle J_x\rangle + |\mathcal{E}|, \\
    \langle \dot{B} \rangle &= -\kappa\langle B \rangle - \frac{g}{2}\langle J_y\rangle, \\
    \langle \dot{J_x}\rangle &= -\gamma_2\langle J_x\rangle + 2g\langle E\rangle\langle J_z\rangle, \\
    \langle \dot{J_y}\rangle &= -\gamma_2\langle J_y\rangle - 2g\langle B\rangle\langle J_z\rangle, \\
    \langle\dot{J_z}\rangle &= -\gamma_1(1+\langle J_z\rangle) - 2g\langle E\rangle\langle J_x\rangle + 2g\langle B\rangle\langle J_y\rangle,
\end{align}
where we have additionally assumed long $T_1$, such that $\gamma_2+\gamma_1/2\approx\gamma_2$, and defined real-valued electric and magnetic cavity field observables, $E=(a+a^\dagger)/2$ and $B=(a-a^\dagger)/(2i)$, and real-valued spin observables, $J_x=J_++J_-$ and $J_y=-i(J_+-J_-)$. As higher order terms are kept, more complex physics are captured. For example, the second order equations, MBE2, capture spin-cavity correlations and cavity-mediated spin-spin correlations, while the MBE3 equations further capture interference effects. As additional terms are kept, the dynamics asymptotically approaches that of the Dicke model \cite{carollo_exactness_2021}. The validity of dropping terms in the cumulant expansion depends on the relative strengths of $g_\text{eff}$, $\kappa$ and $\gamma_2$.

\subsection{Adiabatic Elimination}
\label{sec:adiabaticelimination}
A further simplification of the MBE1s may be made when the ratio $g/\kappa$ is small. 
This \textit{bad-cavity} limit assumes no photon memory and that spin evolution does not change the cavity state.
Thus, the cavity may be adiabatically eliminated and treated as a classical back-action field \cite{ansel_optimal_2018}. Setting the time-derivatives of $\langle E\rangle$ and $\langle B\rangle$ to zero and substituting the result into the spin evolution equations yields the classical back-action equations:
\begin{align}
    \langle \dot{J}_x \rangle &= \frac{g^2}{\kappa}\langle J_x\rangle\langle J_z\rangle + \frac{2 g|\mathcal{E}|}{\kappa}\langle J_z\rangle - \gamma_2\langle J_x\rangle, \\
    \langle \dot{J}_y \rangle &= \frac{g^2}{\kappa}\langle J_y\rangle\langle J_z\rangle - \gamma_2 \langle J_y\rangle\,,\\
    \langle \dot{J}_z \rangle &= -\frac{g^2}{\kappa}(\langle J_y\rangle^2 + \langle J_x\rangle^2) - \frac{2 g|\mathcal{E}|}{\kappa}\langle J_x\rangle - \gamma_1(\langle J_z\rangle + (1 - 2p)),
\end{align}
where we have introduced a polarization, $p=(e^{\hbar\omega_0/kT}+1)^{-1}$, that defines the thermal equilibrium state we are damping to. These equation are identical to the phenomenological radiation-damping Bloch equations (RDBEs) describing the evolution of classical magnetization vectors under back-action from a Rabi drive applied along the y-axis \cite{sodickson_initiation_1996}:
\begin{align}
    \dot{M_x(t)} &= -\frac{M_x(t)}{T_2} + M_z(t)\left(\omega_1-\frac{M_x(t)}{\tau_r}\right),\\
     \dot{M_y(t)} &= -\frac{M_y(t)}{T_2}-\frac{M_y(t)M_z(t)}{\tau_r},\\
      \dot{M_z(t)} &= \frac{1-M_z(t)}{T_1}+\frac{M_y^2(t)}{\tau_r}-M_x(t)\left(\omega_1-\frac{M_x(t)}{\tau_r}\right).
\end{align}
Here we have defined a reduced time-dependent Rabi drive that depends on the back-action:
\begin{equation}
    \omega_1^r=\omega_1-\frac{M_x(t)}{\tau_r}.
\end{equation}
The equivalence between the MBE1s and the RDBEs is formally made with the associations,
\begin{align}
    &\tau_r=(2\pi\eta M_0 Q\gamma)^{-1}\rightarrow\frac{\kappa}{g^2}, \\
    &\omega_1 \rightarrow \frac{2g|\mathcal{E}|}{\kappa},
\end{align}
and a redefinition of the cavity frequency and dissipation rate due to the presence of the spins \cite{abe_electron_2011,diniz_strongly_2011,kurucz_spectroscopic_2011}. For spins resonant with the cavity, this simplifies to:

\begin{align}
    \omega & \rightarrow \omega-g^2\Delta_s/(\Delta_s^2+\gamma_s^2)=0 \\
    \kappa &\rightarrow\kappa +g^2\gamma_2/(\Delta_s^2+\gamma_2^2)=\kappa+g^2/\gamma_2
\end{align}

\section{Transient Measurement of Spin-Cavity Dynamics}

\subsection{Experimental Setup and Characterization}
\begin{figure}[h!]
\centering
\includegraphics[width=1.0\linewidth]{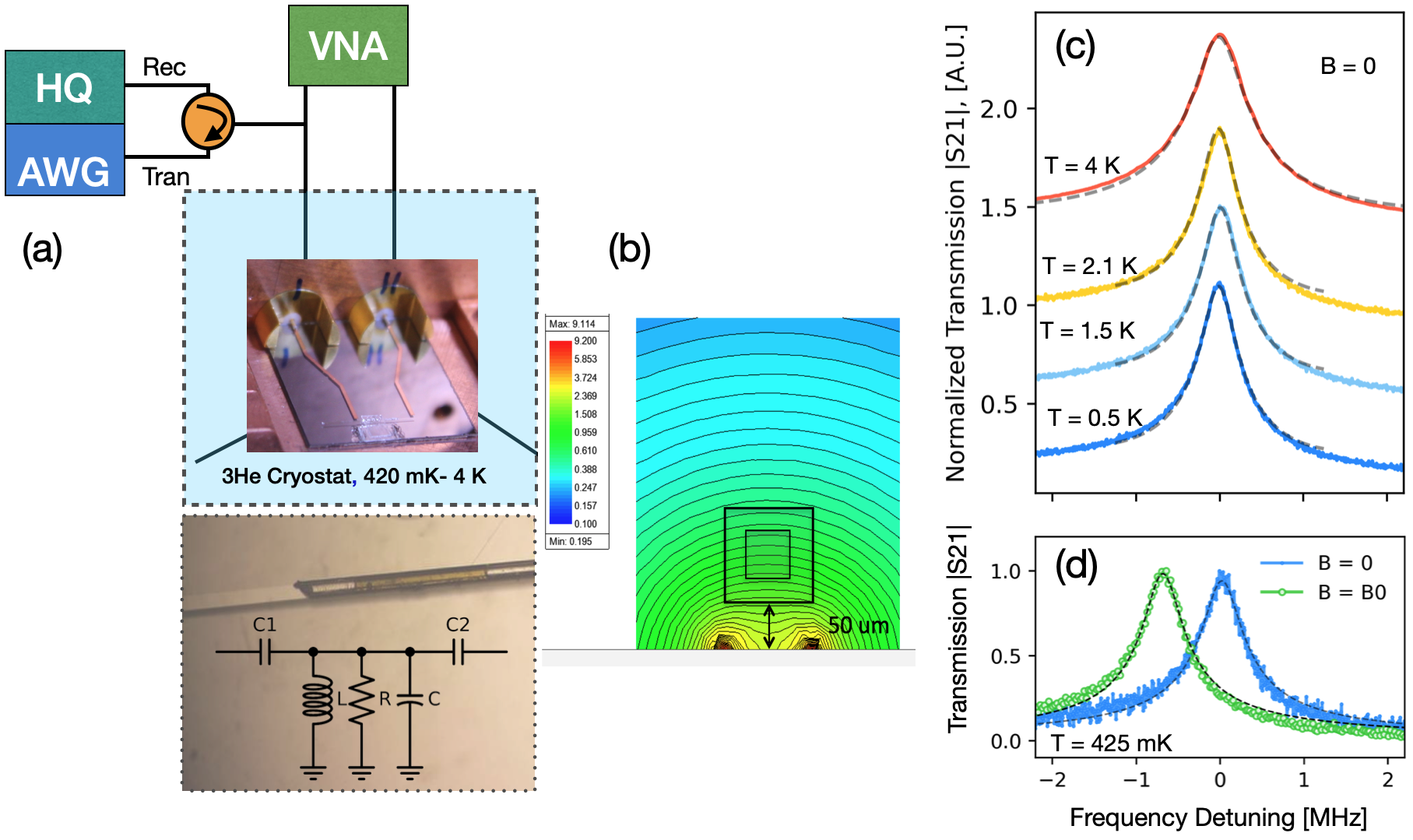}
\caption{Measurement setup schematic and device characterization. \textbf{(a)} A photograph of the device with a 2.5 nL frozen solution of 20 mM Trityl OX63 contained in a capillary mounted approximately 50 $\mu$m above the surface of a capacitively coupled $\lambda$/2 Nb microstrip superconducting resonator patterned on a sapphire chip. The device may be modeled as a two-port RLC resonator with capacitive coupling gaps to microstrip copper feedlines given by $C_1$ and $C_2$. Measurements were performed in a reflection mode with one port terminated in 50$\Omega$. \textbf{(b)} HFSS simulations of the spatial dependence of the microwave field generated by the resonator. A sample height of approximately 75 - 125 $\mu$m was chosen to yield high homogeneity to isolate spin-cavity dynamics. In the sample volume, $g_0$ varies from 0.7 - 1.1 Hz, yielding an expected $g_\text{eff} = \overline{g_0}\sqrt{N}=2.85$ MHz for the $10^{13}$ spins in the sample. \textbf{(c)} Temperature dependence of the resonance at zero field characterized through VNA transmission ($S_{21}$) response. Data is normalized and offset for visual clarity. The device parameters stabilize below 2K, indicating overcoupling below this temperature with resonance frequency $\omega_0=2\pi 9.512$ GHz, external loss $\kappa_e=2\pi 169$ kHz, internal loss $\kappa_i=2\pi 225$ kHz, and total Q = 24,000. \textbf{(d)} Resonator $S_{21}$ response in the presence of a magnetic field of 340 mT. The device was aligned with the external field by rotating the cryostat relative to a room-temperature electromagnet to minimize losses induced by vortex formation and motion. In the final arrangement, Q is unchanged and the resonance shifts approximately 700 kHz below the zero field value.}
\label{fig1}
\end{figure}

A schematic of the experimental set-up is shown in figure \ref{fig1}a. The sample was located in a homogeneous portion of the resonator microwave field 75 - 125 $\mu$m above the resonator surface to simplify the spin dynamics (figure \ref{fig1}b). The superconducting resonator and sample were contained within an oxygen-free copper package and integrated into a custom-built \textsuperscript{3}He cryostat (see SuppMat). Electromagnetic coupling of the microwave irradiation to the device was achieved through adjustment of capacitive gaps between the Nb resonator and copper microstrip feedlines patterned on the sapphire chip. The resulting simplified circuit model of the device is shown in figure \ref{fig1}a, with $C_1$ and $C_2$ representing the coupling capacitances. A homebuilt custom microwave spectrometer operating at X-band (9.5 GHz) was used to transmit microwave signals to the system and demodulate the resulting signal. The spectrometer is equipped with a Quantum Machines FPGA-based arbitrary waveform generator (AWG) and digitizer that are time-locked to one another with a resolution of 1 ns and analog bandwidth of 500 MHz. 

Although the device shown in figure \ref{fig1} is a two-port device, measurements were made in reflection mode, with one port interfaced with a circulator and the other left was open-circuit. The device was characterized using an Agilent N5230A 20 GHz Vector Network Analyzer (VNA) to determine the temperature, power, and field response of the superconducting resonator. $S_{21}$ data presented in figure \ref{fig1}c show the dependence of the resonance on temperature, with resonator quality factor (Q) gradually increasing with decreasing temperature \cite{kwon_magnetic_2018}. The saturation of the resonator Q with temperature is limited because the device is undercoupled below 2.5 K, with internal losses dominant.

The cryostat was rotated to maximize the Q at the operating field to suppress losses associated with vortex formation and motion \cite{kwon_magnetic_2018}. As shown in figure \ref{fig1}d, the resonance frequency shifts less than 1 MHz and the Q remains unchanged when the aligned field is applied. Lorentzian fits of the final configuration (0.34 T; 425 mK) indicate a resonance at $\omega_0=2\pi$ 9.512 GHz with a Q $\approx$ 24,000 with internal loss rate $\kappa_i =2\pi$ 225 kHz and external loss rate $\kappa_e = 2\pi$ 169 kHz.

\subsection{Transient Spectroscopy}
Characterizing spin-cavity dynamics is normally done by either applying a low-amplitude continuous-wave signal to the cavity and measuring the steady-state long-time response as a parameter (typically field) is varied, or by applying a high-amplitude impulse that initiates a short-time response measured as a fast-decaying transient that lasts roughly 2-3 times $T_2$. The flexibility of our microwave control and detection system (see SuppMat) provides access to a hybrid approach: all system parameters are held constant while a low-amplitude continuous-wave signal is applied to the cavity and the transient response is monitored continuously over long periods (ms) with high-resolution (ns). A discrete set of transient acquisitions may be collected as a function of a variable for spectroscopic purposes. Features of the transient response, including the initial transient and the final approach to saturation provide detailed information about the spin-cavity dynamics and the particular parameter regime being investigated. 

In particular, we found that spin-cavity systems in the radiation damping regime exhibit a unique long-lived plateau transient response that strongly depends on $g$, $\kappa$, microwave power, and spin relaxation. The physical origin of the long-lived transient lies in a competition between the action induced by the microwave drive, which rotates the spin magnetization away from thermal equilibrium (z-axis), and the action induced by the back-action field, which appears as a state-dependent drive field shifted 90 degrees from the Rabi drive that tends to rotate the spin magnetization back toward the z-axis \cite{sodickson_initiation_1996,vlassenbroek_radiation_1995}. When these two actions are balanced, a pseudo-steady state is reached that results in a spin response with a lifetime significantly longer than $T_2$. %

\begin{figure}[h!]
\centering
\includegraphics[width=1.0\linewidth]{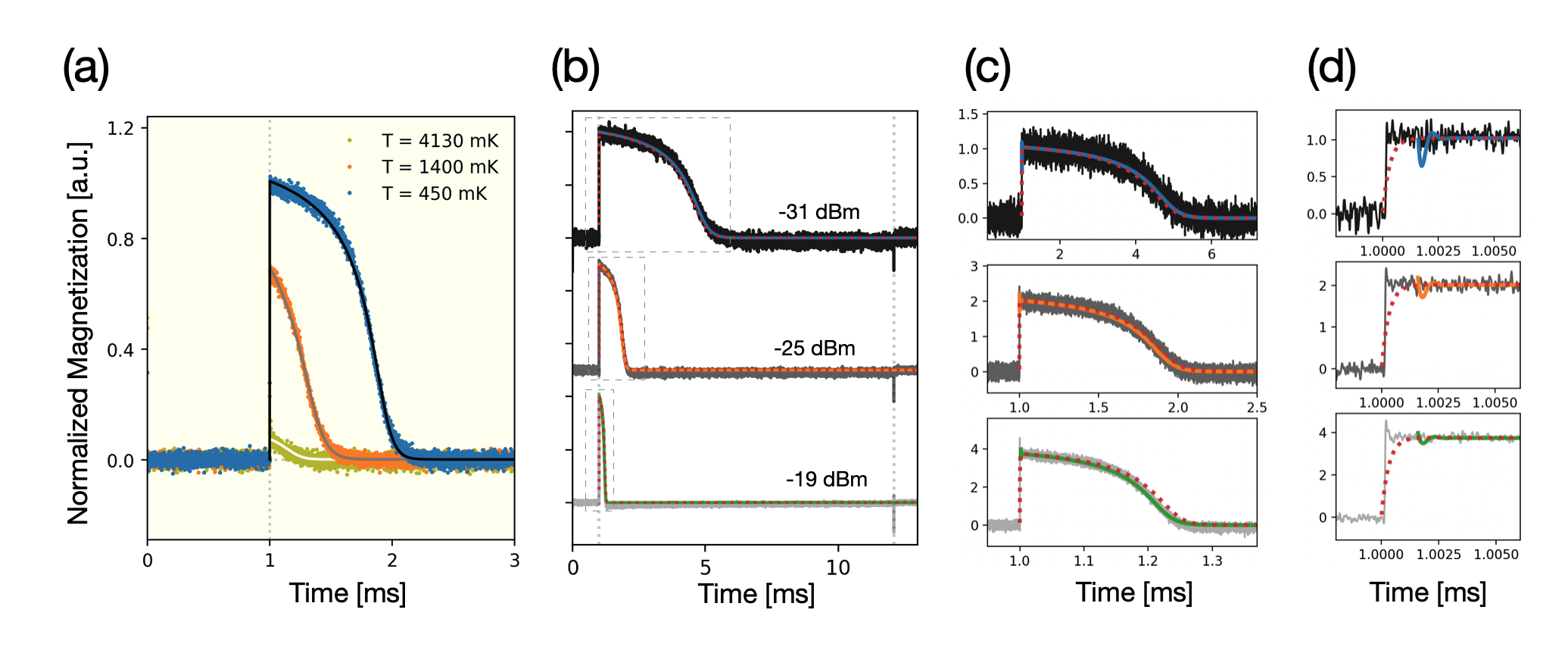}
\caption{Long-lived transient response as a function of temperature and microwave power. \textbf{(a)} Transient response at three different temperatures. A pulse length of 5 ms was used with a microwave power of -25 dBm; however, for clarity, only the first 3 ms are shown. As temperature is lowered, the strength of the spin-cavity coupling, $g$, increases, leading to a more prevalent plateau transient response. The experimental data are overlaid with numerical model fits using the Bloch equations with radiation damping (RDBEs), with fixed parameters $T_1>$ 1 sec, $T_2 = 1.5\mu$s, and $\omega_1=2\pi$ 17.5 kHz. The resulting fits of the radiation damping time constant $\tau_r$ are 1.6 $\mu$s, 1.1 $\mu$s, and 400 ns, respectively, for decreasing temperature. The data for 4.13 K contains 400 averages, while the data for 1.4 K and 0.45 K each contain 10 averages. The data are normalized relative to the maximum signal at the lowest power. \textbf{(b)} Transient response at 425 mK for varying microwave amplitude of a 12 ms pulse. As expected, the length of the plateau response decreases with increased microwave power. MBE model fits yield parameters of $g=2\pi$ 4.2 MHz, $\kappa=2\pi$ 460 kHz, and $\gamma=2\pi$ 670 kHz. \textbf{(c)} Zoomed detail of the microwave power-dependent transient response with model fits using RDBEs (dashed) and MBEs (solid) displayed. The MBE fits capture the final approach to saturation more accurately, indicating a slight violation of the adiabatic elimination approximation in our measurements. \textbf{(d)} Zoomed detail of the microwave power-dependent initial transient with model fits using RDBEs (dashed) and MBEs (solid). Neither model fully captures the details of the initial transient, where spin-cavity correlations are present and have not yet decayed. Higher order MBEs are likely needed to capture this short-time response.}
\label{fig2}
\end{figure}

Initially, we examined the resonant ($\Delta_c=\Delta_s=0$) transient response as a function of temperature. As shown in figure \ref{fig2}a, the prevalence of the plateau transient increases with decreasing temperature, extending for several ms at temperature below 500 mK. Variation of temperature affects the system parameters in multiple ways: spin polarization changes, following the Boltzmann distribution $1-\text{exp}(-\Delta E/(k_B T))$, and modifying $g$; the Q of the resonator changes, modifying $\kappa$; and the spin relaxation times, $T_1$ and $T_2$, change, modifying $\gamma_1$ and $\gamma_2$.

Transient response was also examined as a function of microwave drive amplitude. As shown in figure \ref{fig2}b, the length of the plateau transient response depends inversely on the microwave drive amplitude. Comparisons of model fits using RDBEs versus MBE1s are also shown in figure \ref{fig2}b. As discussed in section \ref{sec:adiabaticelimination}, if the adiabatic elimination conditions hold, the two models should yield identical results. However, the MBE1 fits capture more of the subtle detail present in the long-time approach to saturation, indicating our system is in the strong radiation damping regime where the adiabatic elimination condition is nearly violated. Additionally, closer examination of the initial transient response (figure \ref{fig2}b) indicates neither model captures the short-time behavior where spin-cavity correlations are present and have not yet decayed away. Higher order MBE models are likely necessary to capture these dynamics. The fitted parameters are $\kappa/2\pi=460$ kHz, $\gamma_2/2\pi=670$ kHz, and $g/2\pi=4.2$ MHz. As shown in the Supplementary Material, HFSS field simulations yield an expected $g/2\pi=2.85$ MHz, and steady-state anti-crossing data yield a measured $g/2\pi=4.5$ MHz, in reasonable agreement with the transient spectroscopy fits.

\section{Control of Engineered Spin-Cavity Dynamics}

The degree to which a spin-cavity system exhibits correlated behavior is often quantified using a cooperativity expression \cite{schuster_high-cooperativity_2010}:
\begin{equation}
    C = \frac{g^2}{\kappa\gamma_2}.
\end{equation}
When $C \ll 1$, there is no back-action and the system is in the regime of conventional spectroscopy where only the spin dynamics need be considered. When $C \approx 1-10$, back-action becomes important but may be treated as a mean-field without coherence. This is the regime of radiation-damping and MBE1, with the finer details of the effect of back-action depending on the exact value of $C$, dictating whether the RDBEs or the MBE1s should be used. When $C\gg1$, the system has entered the regime of coherent multi-photon processes and resolved vacuum Rabi splitting between the spins and cavity. In this regime, higher order MBEs must be used that account for the development of spin-cavity correlated states and interference. In situations where the photon occupation is guaranteed to be low, treatment with the Lindblad master equation is feasible. The three regimes of spin-cavity dynamics considered display unique behavior that is appropriate for different applications. The ability to move between the regimes between or during experiments provides powerful flexibility in manipulating spin-cavity systems.

From the standpoint of engineering the spin-cavity system, multiple design considerations dictate which regime the system will naturally operate in. The goal is to design $g_\text{eff}$, $\kappa$, and $T_2$ to yield a target initial cooperativity. The first design parameter we consider is temperature. As seen in figure \ref{fig1}c, the quality factor of the cavity depends on temperature, providing a method to modify $\kappa$. Additionally, the quality factor of the cavity itself may be modified through material choice \cite{zmuidzinas_superconducting_2012} and electromagnetic coupling \cite{rinard_relative_1994} to adjust $\kappa$. Although there is less flexibility in adjusting $T_2$, adjusting the concentration of the spin ensemble is useful. There are several methods to adjust the value of $g_\text{eff}$. The mode volume of the cavity may be adjusted or the sample may be placed closer or further from the field maximum of the cavity (figure \ref{fig1}b) to adjust $g_0$ and its distribution (homogeneity). Additionally, changing the number of spins modifies the value of $g_\text{eff}$ through its dependence on $\sqrt{N}$. 

Aside from hardware engineering of the spin-cavity dynamics, there are several methods to both control the dynamics in a given regime and tune the cooperativity to dictate the operation regime \cite{liensberger_tunable_2021}. Quantum control of state transfer and the generation of a target Liouvillian operation (generalizing beyond unitary dynamics) is well-studied, with many examples and algorithms given in the literature (c.f. \cite{khaneja_optimal_2003,haeberlen_coherent_1968,borneman_application_2010,borneman_bandwidth-limited_2012,endeward_implementation_2023,motzoi_simple_2009}). Additionally, there are several examples of generating control sequences that use both the RDBEs and MBEs \cite{ansel_optimal_2018,krimer_optimal_2017,zhang_time-optimal_2011}. It has also been noted and demonstrated that adjusting the magnetic field to move the spins into and out of resonance with the cavity, either adiabatically or diabatically, permits an effective switch of the spin-cavity coupling Hamiltonian, which becomes non-secular and suppressed when the field is detuned \cite{albanese_radiative_2020,wood_cavity_2014}. 

\begin{figure}[h!]
\centering
\includegraphics[width=0.8\linewidth]{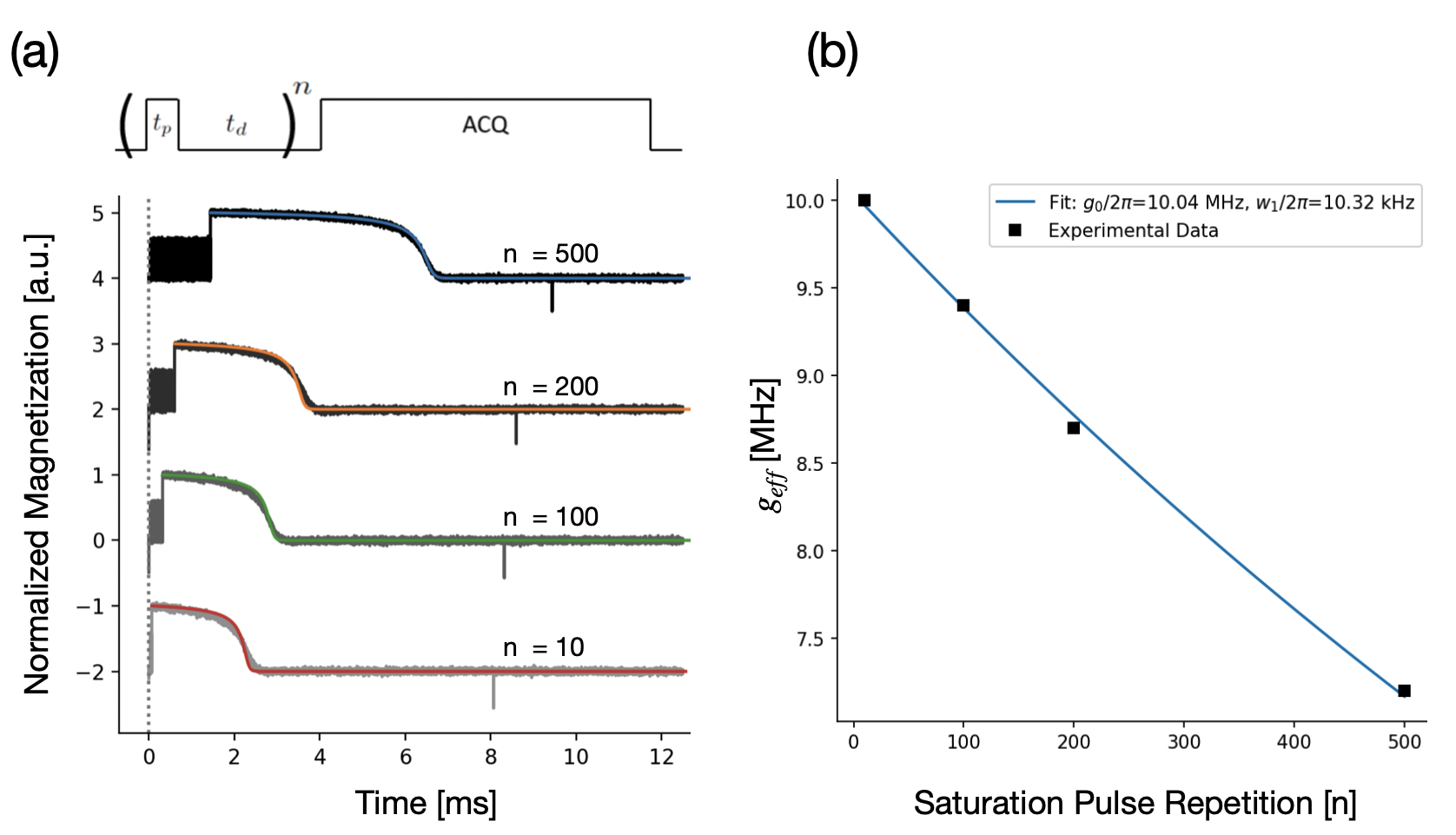}
\caption{Long-lived transient response as a function of presaturation. Prior to acquisition of the spin-cavity transient response, a series of pulses of length $t_p$ followed by a delay of length $t_d$ are repeated $n$ times. The excited magnetization decoheres during $t_d>>T_2$, such that the excited spins do not contribute to the dynamics. As the number of repetitions of the saturation sequence is increased, more spins are saturated, reducing $N_\text{eff}$ and reducing $g_\text{eff}$. \textbf{(a)} Transient spectroscopy measurements after saturation sequences of $n=10,100,200,500$. This measurement was performed on a 100 mM Trityl OX63 sample at T = 600 mK, using $t_p=800$ ns, $t_d=2$ $\mu$s, and $\omega_1/2\pi=10.3$ kHz. The lifetime of the plateau response varies inversely with $g_\text{eff}$
, providing a measure of the effective number of spins contributing to the spin-cavity dynamics through $g_\text{eff}=g_0\sqrt{N_\text{eff}}$. 
In this measurement, 10 repetitions yields $g_\text{eff}=2\pi$ 10.0 MHz, 100 repetitions yields $g_\text{eff}=2\pi$ 9.4 MHz, 200 repetitions yields $g_\text{eff}=2\pi$ 8.7 MHz, and 500 repetitions yields $g_\text{eff}=2\pi$ 7.2 MHz. \textbf{(b)} Comparison of experimentally determined $g_\text{eff}(n)$ (black squares) to the theoretical model of $g_\text{eff}(n)=g_\text{eff}(0)\sqrt{(\cos{\omega_1t_p})^n}$ } 
\label{fig3}
\end{figure}

Direct modulation of $g_0\rightarrow g_0(t)$ has been demonstrated and provides a powerful control of spin-cavity dynamics \cite{sung_realization_2021}. However, the range of coupling strengths that may be accessed is limited. An alternative method to tuning the cooperativity is to change the number of spins that effectively contribute to the dynamics, thereby changing $g_\text{eff}$ \cite{rose_coherent_2017,putz_protecting_2014}. As shown in figure \ref{fig3}, a progressive saturation sequence may be used to set a certain number of spins to an Identity state. Changing the degree of saturation changes the behavior of the plateau transient response, indicating a modification of $g_\text{eff}$ through a change in the number of spins. The dependence of $g_\text{eff}$ on the number of small-angle saturation pulses applied is given by
\begin{equation}
    g(n)=g(0)\sqrt{(\cos{\omega_1t_p})^n}.
\end{equation}
As shown in Fig. \ref{fig3}b, a fit to the experimental data with $\omega_1/2\pi=10.3$ kHz explains the observed variation in $g_\text{eff}$ for $t_p=800$ ns and $t_d\gg T_2$.

\section{Discussion}

We have demonstrated a hybrid spin-cavity system that admits engineering of the parameters that uniquely define various regimes of correlation: photon exchange rate, $g_\text{eff}$, and cavity dissipation rate, $\kappa$. A combination of changing sample placement, ensemble size, resonator Q, resonator coupling, and temperature enables tuning the system into a desired parameter regime that exhibits target dynamics, including no back-action, classical back-action (radiation damping), and correlated behavior. The dynamics may be modeled by Maxwell-Bloch equations that are shown to agree with radiation damping Bloch equations models. The transient behavior was measured as a function of temperature, microwave drive power, and spin saturation, providing a suite of controls to tune the cooperativity of the device.

We have also demonstrated a new method to characterize spin-cavity dynamics through low-amplitude continuous-wave transient detection. In the presence of cavity back-action, the transient response of the system is characterized by a long-lived plateau that provides detailed information about the subtle differences in different spin-cavity models. Importantly, this method provides a strong, useful signal even in the presence of long sample $T_1$, which is characteristic of low-temperature systems used for quantum devices. The utility of this method for general spectroscopy needs to be studied in more detail. Additionally, due to the low microwave drive amplitudes involved in this method, it is compatible with quantum-limited amplifiers, including Josephson parameteric amplifiers (JPAs), providing a path toward quantum-limited detection methods of correlated spin-cavity dynamics \cite{bienfait_reaching_2016,travesedo_all-microwave_2025}. Future studies will also include further examination of multiphoton processes that violate the approximations of the MBE1 model and the development of new models that explicitly include the detection system to account for behavior noted in radiation damping literature, such as the use of Q-spoiling or coherent feedback to suppress back-action \cite{broekaert_suppression_1995,maas_suppression_1995,anklin_probehead_1995}.

\section{Acknowledgements}
This research was undertaken thanks in part to funding from the Canada First Research Excellence Fund. We thank Dr. Dmitry Akhmetzyanov for helpful discussions on ESR experiments and sample preparation, and Ruhi Shah for insightful discussions on the derivation of the Maxwell-Bloch equations.

\section{Author Contributions}
F.F. prepared the sample, performed the experiments, and analyzed the data. G.N. constructed the \textsuperscript{3}He experimental apparatus. I.T. designed and assembled the microwave package and mounted the device and sample. H.M. designed the superconducting resonator. S.S. fabricated the superconducting resonator (KC006248). T.B. and D.C. conceived and supervised the project. F.F. and T.B. wrote the paper with input from all authors.

\section{References}

\bibliographystyle{unsrt}
\bibliography{spincavityrefs.bib}

\section{Figures and Tables}

\newpage

\newpage

\end{document}